\documentstyle[emulateapj]{article}

\begin{document}
\title{DISCOVERY OF THE OPTICAL TRANSIENT OF THE GAMMA RAY BURST 990308}

\author{Bradley E. Schaefer$^{a,b}$, 
J. A. Snyder$^{a}$,
J. Hernandez$^{c,d}$, 
B. Roscherr$^{a}$, 
M. Deng$^{a}$, 
N. Ellman$^{a}$, 
C. Bailyn$^{b}$, 
A. Rengstorf$^{e}$, 
D. Smith$^{f}$, 
A. Levine$^{f}$, 
S. Barthelmy$^{g}$, 
P. Butterworth$^{g}$, 
K. Hurley$^{h}$, 
T. Cline$^{g}$, 
C. Meegan$^{i}$, 
C. Kouveliotou$^{i}$, 
R. M. Kippen$^{i}$,  
H.-S. Park$^{j}$, 
G. G. Williams$^{k}$, 
R. Porrata$^{j}$, 
R. Bionta$^{j}$, 
D. Hartmann$^{k}$, 
D. Band$^{l}$, 
D. Frail$^{m}$, 
S. Kulkarni$^{n}$, 
J. Bloom$^{n}$, 
S. Djorgovski$^{n}$,
D. Sadava$^{n}$,
F. Chaffee$^{p}$, 
F. Harris$^{q}$, 
C. Abad$^{c}$, 
B. Adams$^{e}$, 
P. Andrews$^{a}$, 
C. Baltay$^{a}$, 
A. Bongiovanni$^{c, r}$, 
C. Briceno$^{a}$
G. Bruzual$^{c}$, 
P. Coppi$^{a,b}$,
F. Della Prugna$^{c}$, 
A. Dubuc$^{c}$, 
W. Emmet$^{a}$, 
I. Ferrin$^{d}$, 
F. Fuenmayor$^{d}$, 
M. Gebhard$^{e}$, 
D. Herrera$^{c, d}$, 
K. Honeycutt$^{e}$, 
G. Magris$^{c}$, 
J. Mateu$^{c, d}$, 
S. Muffson$^{e}$, 
J. Musser$^{e}$, 
O. Naranjo$^{d}$, 
A. Oemler$^{b,o}$, 
R. Pacheco$^{c, d}$, 
G. Paredes$^{c, d}$, 
M. Rengel$^{c, d}$, 
L. Romero$^{c, d}$, 
P. Rosenzweig$^{d}$, 
C. Sabbey$^{b}$, 
Ge. S\'{a}nchez$^{c}$,  
Gu. S\'{a}nchez$^{c}$, 
H. Schenner$^{c}$, 
J. Shin$^{a}$,
J. Sinnott$^{a}$, 
S. Sofia$^{b}$, 
J. Stock$^{c}$,
J. Suarez$^{d}$,
D. Tell\'{e}ria$^{c,d}$, 
B. Vicente$^{c}$, 
K. Vieira$^{c,d}$, 
K. Vivas$^{b}$
}

\begin{center}
$^{a}$Yale University, Physics Department, P. O. Box 208121, New Haven CT
06520-8121

$^{b}$Yale University, Astronomy Department, 260 Whitney, New
Haven CT 06511

$^{c}$Centro de Investigaciones de Astronom\'{i}a (CIDA), A. P.
264, M\'{e}rida 5101-A, Venezuela

$^{d}$Universidad de los Andes, Departamento de Fisica, 5101, M\'{e}rida,
Venezuela

$^{e}$Indiana University, Dept. of Astronomy, 319 Swain West, Bloomington
IN, 47405

$^{f}$Massachusetts Inst. of Technology, Physics Dept., 77 Mass. Ave,
Cambridge MA 02138

$^{g}$NASA Goddard Space Flight Center, Code 660, Greenbelt MD 20771

$^{h}$University of California, Space Science Lab., Berkeley CA 94720-7450

$^{i}$NASA Marshall Space Flight Center, ES-84, Huntsville AL, 35812

$^{j}$Lawrence Livermore National Lab., PO Box 808, MS L274, Livermore CA
94550

$^{k}$Clemson University, Dept. of Physics \& Astronomy, Kinard Lab.,
Clemson SC, 29634

$^{l}$Los Alamos National Lab., MS D436, Los Alamos NM, 87545

$^{m}$National Radio Astronomy Observatory, P. O. Box O, Socorro, NM 87801

$^{n}$California Institute of Technology, Astronomy, MS 105-24, Pasadena
CA, 91125

$^{o}$Carnegie Observatories, 813 Santa Barbara St., Pasadena CA, 91101

$^{p}$W. M. Keck Observatory, 65-0120 Mamalahoa Highway, Kamuela HI,
96743

$^{q}$U. S. Naval Observatory, P. O. Box 1149, Flagstaff AZ, 86002

$^{r}$Universidad Central de Venezuela, Departamento de Fisica, 1042
Caracas, Venezuela

\end{center}

\begin{abstract}

The optical transient of the faint Gamma Ray Burst 990308 was detected by the 
QUEST camera on the Venezuelan 1-m Schmidt telescope starting 3.28 hours
after the burst. Our photometry gives $V = 18.32 \pm 0.07$, $R = 18.14 \pm 
0.06$, $B = 18.65 \pm 0.23$, and $R = 18.22 \pm 0.05$ for times ranging
from 3.28 to 3.47 hours after the burst. The colors correspond to a
spectral slope of close to $f_{\nu} \propto \nu^{1/3}$.  Within the
standard synchrotron fireball model, this requires that the external medium be 
less dense than $10^{4} cm^{-3}$, the electrons contain $> 20\%$ of the
shock energy, and the magnetic field energy must be less than $24 \%$ of 
the energy in the electrons for normal interstellar or circumstellar
densities. We also report upper limits of $V > 12.0$ at 132 s (with
LOTIS), $V > 13.4$ from 132-1029s (with LOTIS), $V > 15.3$ at 28.2 min 
(with Super-LOTIS), and a 8.5 GHz flux of $< 114 \mu Jy$ at 110 days (with
the Very Large Array). WIYN 3.5-m and Keck 10-m telescopes reveal this
location to be empty of any host galaxy to $R > 25.7$ and $K > 23.3$.  
The lack of a host galaxy likely implies that it is either substantially 
subluminous or more distant than a red shift of $\sim 1.2$.

\end{abstract}


\section{INTRODUCTION}

The old dilemma of whether Gamma Ray Bursts (GRBs) are galactic or cosmological 
has been solved recently with the discovery of X-ray (Costa et al. 1997), 
optical (van Paradijs et al. 1997), and radio transients (Frail et al.
1998).  
Four optical transients have measured red shifts of 0.835 (GRB970508), 0.966 
(GRB980703), $\geq 1.600$ (GRB990123), and $\geq 1.619$ (GRB990510), thus
proving that bursts are at cosmological distances. 
 
After the burst is over, the expanding shell emits an afterglow which 
eventually fades to invisibility.  Early detections of the afterglow include 
 GRB990510 at 3.52 hours (Harrison et al. 1999) and GRB990123
during the burst by ROTSE (Akerlof et al. 1999). Theoretical models 
of synchrotron emission from a decelerating 
relativistic shell shocked by collision with an external medium 
are in good agreement with afterglow observations from X-ray 
to radio and from the earliest to the latest times of available observations 
(Galama et al. 1998; 1999). 

The accurate optical/radio positions allow for 
very deep searches after the transient fades. For the 11  
optical/radio transient positions searched to date, a very faint galaxy 
(typically with $R \sim 25$) appears within an arc-second or so for 9 of
the positions. This
result demonstrates that most GRBs reside inside distant host galaxies.
Although the optical/radio transient positions are within the galaxies, they are significantly 
offset from the centers, demonstrating that the progenitors are neither 
associated with a giant central black hole nor are ejected from the galaxy. 
Bursts have been associated with active star formation (Totani 1997,
 Paczynski 1998), while the hosts have been shown to be significantly 
subluminous for normal galaxies in the majority of cases (Schaefer 1999). 
Three of the positionally coincident host galaxies have observed redshift 
values of 0.695 (GRB970228), 3.418 (GRB971214), and 1.097 (GRB980613), in 
reasonable agreement with the redshifts from the optical transients.

The GRB explosion mechanism is still unknown, but a profitable line of 
study is to observe the burst afterglows over a wide range of time and 
frequency, with demographic arguments providing information on the progenitor 
and hence the explosion mechanism. Few afterglows have been observed to date, 
so further examples are important. This paper reports the discovery of the 
afterglow of GRB990308 with B, V, and R imagery starting 3.28 hours after the 
burst.

\section{OBSERVATIONS}

GRB990308 triggered the Compton Gamma Ray Observatory Burst and 
Transient Source Experiment (BATSE) on 1999 March 8 05:15:07 UT. 
This burst has a peak flux (50-300 keV, 256 ms time bins) of 
$1.6 \pm 0.1 ph \cdot s^{-1} \cdot cm^{-2}$, a fluence ($> 25$ keV) of
$2.2 \times 10^{-5} erg \cdot cm^{-2}$, and a $T_{90}$ duration of $106
\pm 12$ s. GRB990308 also triggered one camera of the Rossi X-Ray Timing
Explorer (RXTE) All-Sky Monitor (Levine et al. 1996; 
Smith et al. 1999a).  At 06:32 UT, a preliminary position had been determined 
and distributed as e-mail by the GRB Coordinate Network (GCN; Barthelmy et al. 
1994).  This position is a long thin error box which has since been updated 
(see below) to one with a $3-\sigma$ total width of close to 6.1'. 

GRB990308 was also detected,
very weakly, by the Ulysses GRB detector (Hurley et al. 1992). A cross 
correlation of the light curves between Ulysses and BATSE (separated by 
2066.004 light seconds) as part of the Interplanetary Network (IPN; Hurley 
et al. 1999) yields an annulus with $3-\sigma$ 
total-width of 35.9'. The IPN/RXTE error boxes overlap at an angle of 
$\sim 80^{\circ}$ to form
a $3-\sigma$ error box defined by four points with J2000 coordinates of
12:22:12 +06:35:06, 12:22:23 +06:29:06, 12:24:05 +06:57:40, and 12:24:18
+06:52:05.  

The final error region was imaged with the LOTIS (Park et al. 1997) and Super-
LOTIS (Park et al. 1998) wide-field fast response cameras located at the 
Lawrence Livermore National Laboratory.  Both systems responded to a GCN 
notice of the preliminary BATSE coordinates, and the first images were taken 
12.6 s after the trigger.  An updated BATSE position from the GCN 
shifted the position, with the cameras responding quickly, such that the first 
LOTIS images that cover the error box are at 132 s after the BATSE 
trigger. This 10 s integration reached  $V = 12.0$ (S/N 
= 3, no filter) with no indication of an optical transient.  A sum of
roughly ten minutes of exposure from 132-1029 s after the BATSE trigger 
showed no optical transient to $V = 13.4$.  Unfortunately, the night was
somewhat foggy, so the limiting magnitude was dominated by the unusually
bright sky background.  The Super-LOTIS camera has a $0.8^{\circ}$ square
field-of-view which took 30 s exposures in a spiral pattern around the 
center of the BATSE error region.  The burst position was imaged starting 
28.2 min after the burst to a depth of $V = 15.3$ (S/N = 3, no filter)
with no evidence for an optical transient.
 
With the QUEST camera on the 1.0-m Schmidt telescope of the Observatorio
Nacional de Llano del Hato near M\'{e}rida, Venezuela, we started scans of the 
preliminary error box approximately three hours after the burst.  The QUEST 
camera consists of sixteen $2048 \times 2048$ CCD detectors in a $4
\times 4$ array operating in a drift scan mode (Snyder 1998, Sabbey,
Coppi, \& Oemler 1998).  At any instant the QUEST camera is viewing 5.4 square 
degrees and in one hour will scan a $2.3^{\circ} \times 15^{\circ}$ 
region, which makes the camera ideal for rapid response to large preliminary 
GRB error boxes.  Each row of detectors is covered with a different filter, so 
that the GRB990308 error box was imaged four independent times with standard 
(Bessel 1990) V, R, B, and R bands in time order.  The integration time for 
each image was 142 s, with a total time of 11.5 min separating the centers
of first and last. The limiting magnitudes were $B = 19.2, V = 20.2$, and
$R = 21.6$ to the $S/N = 3$ level with the 69\% illuminated Moon
$53^{\circ}$ away. The pixel size is 1.02'', the FWHM seeing was 2.6'', 
and the entire IPN/RXTE box was covered. 

The large size of the QUEST field-of-view is good for rapid imaging of large 
preliminary GRB positions, but it also means that rapid identification is 
difficult until a small error box is reported.  For GRB990308, the early error 
box was $8.6^{\circ}$ long until March 14 when the small IPN/RXTE region was 
produced (Smith et al. 1999b).  The deep limiting magnitude of QUEST is good 
for catching faint optical transients, but it also means that the afterglow 
cannot be recognized if it is near or below the threshold of the Digital Sky 
Survey until deeper comparison images can be made.  For GRB990308 
where our combined images go to $R = 22.1$ for $S/N = 3$ with no obvious
transient brighter than $R = 18$, we had to await the acquisition of deep
comparison images. These were made with the Yale 1-m telescope at Cerro Tololo 
(operated by the YALO consortium; Mendez, Depoy, and Bailyn 1998) in the 
R-band on 1999 May 29 as a 14 element mosaic reaching $R = 22.9$ mag.
A star-by-star comparison rapidly revealed an optical transient close to the 
center of the IPN/RXTE region (Fig. 1). 

The transient appears with a S/N of 15.0, 17.1, 4.7, and 20.8 on the V,
R, B, and R images, respectively.  All four images show the transient with
the 
shape and FWHM of our point-spread-function.  The astrometric position on
all four images is identical to within an rms scatter of 0.4'', with the fitted
 motion during the 11.5 min of observation equal to $0.03''\pm 0.28''$. We
take the appearance of four independent and significant images with good 
stellar shapes as proof that the images are not artifacts of any kind.  For 
the position near opposition, the lack of motion rules out all Solar System 
objects, including Kuiper Belt Objects.  The source must have an amplitude of 
$>7.5$ mag since the source position is empty to $R > 25.7$ in June (see
below).  The only known astrophysical objects outside our Solar System with 
such amplitudes are dwarf novae, novae, supernovae, and GRBs.  The transient 
cannot be a dwarf nova or nova since the $R = 18.14$ observed magnitude
(see below) would imply a distance far outside any galaxy. The transient 
cannot be a supernova (with $M_{R} > -19.5$ for a distance modulus of
$< 37.64$) as our limit on any parent galaxy (which must have $R > 25.7$, 
see later)
would then be $M_{R} > -11.94$. The transient is rather similar to GRBs seen 
previously (van Paradijs et al. 1997; Galama et al. 1998; 1999; Harrison
et al. 1999).  Thus we conclude that our optical transient is definitely the 
afterglow of GRB990308.

We calibrated the B, V, and R magnitudes of nearby comparison stars with images 
taken on 1999 June 10 with the Yale 1-m telescope.  This calibration was made 
by the observation of 22 standards stars (Landolt 1992) which was then applied 
to images of the GRB990308 field.  We then performed differential photometry 
on the QUEST images with respect to 4 nearby comparison stars of known 
magnitude.  We find $V = 18.32 \pm 0.07$ at 196.8 min, $R = 18.14 \pm
0.06$ at 200.6 min, $B = 18.65 \pm 0.23$ at 204.5 min, and $R = 18.22
\pm 0.05$ at 208.3 min after the burst.  The times are for the
middle of the integrations relative to the BATSE trigger.  

The position of the transient was measured for all four images with respect to 
stars in the USNO-A2.0 catalog (Monet et al. 1998). The combined position is 
J2000 $12:23:11.44 \pm 0.02$ $+06:44:05.10 \pm 0.17$.  For comparison
purposes, our star \#1 ($B = 14.75 \pm 0.01$, $V = 13.94 \pm 0.02$,
$R = 13.50 \pm 0.03$) is at $12:23:11.272 +06:45:38.19$ while star \#6
($B = 18.40 \pm 0.07$, $V = 17.63 \pm 0.03$, $R = 17.16 \pm 0.03$) is at
12:23:10.824 +06:43:20.67.

We observed this position with the Very Large Array telescope and found no 
significant radio sources to the $3-\sigma$ level of $258 \mu Jy$ at 8.5
GHz on 1999 June 18, to $114 \mu Jy$ at 8.5 GHz on 1999 June 26, and
to $165 \mu Jy$ at 1.4 GHz on 1999 July 4. 

The transient had probably faded to invisibility by the mid-March, but
late time imaging was carried out to find the GRB host galaxy. The transient position 
is empty on the Palomar Sky Survey and our Yale 1-m 
images.  We took deep R-band images with the WIYN 3.5-m on Kitt Peak on 1999 
June 12.  Our six 15-min exposures were processed and co-added by the 
normal procedures.  The resulting picture shows no significant source within 
6'' to a $S/N = 3$ limiting magnitude of $R = 25.4$.  

We also obtained deep K-band and R-band images of the optical transient 
position with the 10-m Keck I and II telescopes, respectively.  The R-band
image was obtained on 1999 June 19 as a combination of five 200 s 
exposures with the Low-Resolution Imaging Spectrometer (Oke et al. 1995). 
This reached a
$S/N = 3$ limiting magnitude of $R = 25.7$. No star or galaxy appears to this 
limit within $6.3''$ of the optical transient position. The K-band images were 
obtained on 1999 June 23 and 24 as a combination of 12 second exposures
totaling 
71 and 72 min with the Near Infrared Camera (Matthews \& Soifer 1994).
The nights were photometric 
and we obtained standard star observations to calibrate the photometry.  The 
fields are roughly 38.4'' on a side with 0.15Ó pixels, reaching a $S/N = 3$
limiting magnitude of $K = 23.3$. The position of the optical transient
was empty to this threshold.

\section{ANALYSIS}

Our observations can be used to constrain the decline rate of the afterglow. 
We will parameterize the decline with a power law index $\delta$, such
that the flux varies as $t^{\delta}$ with $t$ the time since
the burst. The two QUEST R-band observations 7.7 min apart show a fading 
by $0.08 \pm 0.08$ mag which suggests that $\delta = - 2 \pm 2$.  The
QUEST V-band measurements can be combined with each of the three limits from
LOTIS and Super-LOTIS to give similar constraints that $\delta > -1.3$ for
early times.  The WIYN and Keck limits imply that $\delta < -1.1$ for late
times. Taken together, all of our detections and limits are consistent
with $\delta = -1.2 \pm 0.1$ for the assumption of a single power law over
all time. As discussed below, the spectral slope gives reason to expect a more 
complicated time dependance with $\delta$ close to zero for early times.

The QUEST B, V, and R magnitudes can yield a spectral slope.  The first step 
is to correct for the known galactic extinction of E(B-V) = 0.023 (Schlegel, 
Finkbeiner, \& Davis 1998).  The second step is to deduce the magnitudes and 
colors at one instant of time, which we take as the time of our first image. 
If $\delta = -1$ then the colors are $B - V = 0.26 \pm 0.24$ and
$V - R = 0.17 \pm 0.08$, while if $\delta = 0$ then the colors are
$B - V = 0.30 \pm 0.24$ and $V - R = 0.13 \pm 0.08$.  The third step is to
convert these magnitudes into flux units in Janskys, $f_{\nu}$.  The
fourth step is to fit the three fluxes to a presumed power law with 
$f_{\nu} \propto \nu^{\alpha}$.  We find $\alpha = 0.38 \pm 0.25$ for
$\delta = -1$ and $\alpha = 0.49 \pm 0.26$ for $\delta = 0$.

Within the framework of afterglow models for the synchrotron emission from 
external shocks (Sari, Piran, and Narayan 1998), the spectral slope is either 
in an $\alpha = 1/3$ regime or in regimes with $\alpha \leq -0.5$.  Our
measured spectral slope is easily consistent with the $\alpha = 1/3$ regime, 
yet is inconsistent with the $\alpha \leq -0.5$ regime at the $5
\times 10^{-4}$ probability level.  Thus we conclude that the 
afterglow was in the $\nu^{1/3}$ regime at a time 3.28 hours after the burst.

Synchrotron models require that any afterglow in the $\nu^{1/3}$ regime
have $-1/3 \leq \delta \leq 1/2$ (Sari, Piran, and Narayan 1998), which is 
to say 
that the afterglow is not fast-fading and may even be brightening.  GRB970508 
provides a precedent for an afterglow brightening for the first two days,
during 
which time the spectral slope is positive and consistent with $\nu^{1/3}$ 
(Castro-Tirado et al. 1998).  A $\delta \sim 0$ light curve for GRB990308 
would explain the absence of a detection by LOTIS and Super-LOTIS, yet is not 
significantly inconsistent with our observed decline by $0.08 \pm 0.08$ 
mag in 7.7 minutes.

Within the standard synchrotron fireball model, we can place interesting 
constraints on the burst energetics. For the GRB 990308 optical light to
be in the
$\nu^{1/3}$ regime, the characteristic times $t_{c}$ and $t_{m}$  (Sari,
Piran, \& Narayan 1998, eqns 15 and 16) must both be greater 
than 3.28 hours.  For the case of adiabatic evolution, this requires
$E_{52} > 0.0023 \epsilon_{B}^{-1} \epsilon_{e}^{-4}$ and
$E_{52} < 0.0002 \epsilon_{B}^{-3} n^{-2}$, where $E_{52}$ is the energy
in the spherical shock in units of $10^{52}$ ergs, n is the number density in 
the external medium in units of $cm^{-3}$, while $\epsilon_{e}$ and
$\epsilon_{B}$ are the fractions of the shock energy in the electrons and
magnetic field.  These two constraints limit the allowed values to a
small sliver of $E_{52} - \epsilon_{B}$ parameter space, centered roughly on 
the $40\epsilon_{B} = E_{52}^{-0.5}$ line.  With $n \geq 1$ and the
required $\epsilon_{e} \leq 1$, then $\epsilon_{B} \leq 0.25$ and
$E_{52} > 0.01$.  Since for reasonable models and beaming factors we have
$E_{52} < 100$ and $n > 1$, we derive $\epsilon_{e} > 0.2$.  Similarly,
for $E_{52} < 100$ and $\epsilon_{e} \leq 1$, we find that $n < 10^{4}$,
which might be a problem for models where the burster sits in a massive wind 
from a massive supernova precursor or in dense star forming regions. For the 
case of equipartition of energy, there is no acceptable solution other than in 
the case that $n < 0.1$. For the cases of $n = 1$ and $n = 1000$,
$\epsilon_{B}$ must be less than $24\%$ and $0.03\%$ of $\epsilon_{e}$
respectively with both $t_{c}$ and $t_{m}$ pushed to 3.28 hours. Similar
conclusions are reached in the radiative 
case with the Lorentz factor of the shocked material is $> 10$.  So, with
GRB990308 being in the $\nu^{1/3}$ regime at 3.28 hours after the burst
for standard fireball models with $n > 0.1$, we argue that the external
medium has $n < 10^{4}$, that the fraction of the shock energy in electrons 
must be greater than 20\%, that the fraction of the shock energy in the 
magnetic field must be less than 25\%, and that the electrons  not be in 
equipartition with the magnetic field.

It is generally thought that GRBs were born in distant host galaxies, so
the 
absence of any possible host galaxy to $R = 25.7$ is problematic. A
possible solution to the lack of a visible host galaxy is that the 
burster was ejected from its home galaxy long ago.  The possibility of 
ejection is natural in some models, such as neutron star collision scenarios, 
where the progenitor binary system may receive large velocity kicks during 
neutron star formation. This solution might be implausible since all other 
optical/radio transients with exact positions and associated hosts are within 
the visible galaxies, so at least these bursters have not been ejected.  
For the specific case of GRB990308, the nearest galaxy is one 
with $R = 24.7$ at 6.3'' angular distance, which implies a transverse 
separation of $\sim 50$ kpc or more for a plausible host and luminosity.
Such a transverse separation is unlikely within current models (Bloom, 
Sigurdsson, \& Pols 1999), so we consider ejection improbable.  

The second possible solution to the lack of a visible host is that the galaxy 
is substantially subluminous.  This case has already been demonstrated for the 
majority of the very brightest bursts irrespective of the breadth of the GRB 
luminosity function (Schaefer 1999), while several of the hosts associated 
with optical transients with redshifts are in the lower few per cent of the 
luminosity-weighted Schechter luminosity function (GRB970228, GRB970508, 
GRB990510).  Indeed, if GRB990308 has the average luminosity associated with 
the no-evolution fits to the LogN-LogP curve ($10^{57} ph \cdot s^{-1}$ or
$6 \times 10^{50} erg \cdot s^{-1}$; Fenimore et al. 1993), then the host
galaxy would be at $z = 0.50$ with an absolute magnitude fainter than
$-17.2$ which is in the faintest $\sim 2\%$ of galaxies.

The third possible solution to the lack of a visible host is that the burst 
can be at a high redshift.  For an average galaxy, say one in the middle of 
the luminosity-weighted luminosity function with an absolute R magnitude of 
$-20.8$, the burst must have $z > 1.2$ for us to not detect the host to
$R = 25.7$.  Here and throughout this paper, we have adopted a Hubble Constant 
of $65 km \cdot s^{-1} \cdot Mpc^{-1}$, $\Omega_{0} = 0.3$, $\Omega_{\Lambda} 
= 0.7$, the galaxy K-corrections
for Sb galaxies of Rocca-Volmerange \& Guideroni (1988), and the GRB
K-corrections of Fenimore et al. (1992) with spectral slope -1.5.  With this 
approximate limit based on an assumption that the host is average in brightness
, the deduced burst luminosity must be $> 6 \times 10^{57} ph \cdot
s^{-1}$. Pushing GRB990308 to $z > 1.2$ is plausible. 

This work was supported by the National Science Foundation, the Department of 
Energy, and the National Aeronautics and Space Administration, while the
Observatorio Astron\'{o}mico Nacional is operated by
CIDA for the Consejo Nacional de Investigaciones Cient\'{\i}ficas y
Tecnol\'{o}gicas.

\begin{figure}
{\epsscale{0.4}
\plotone{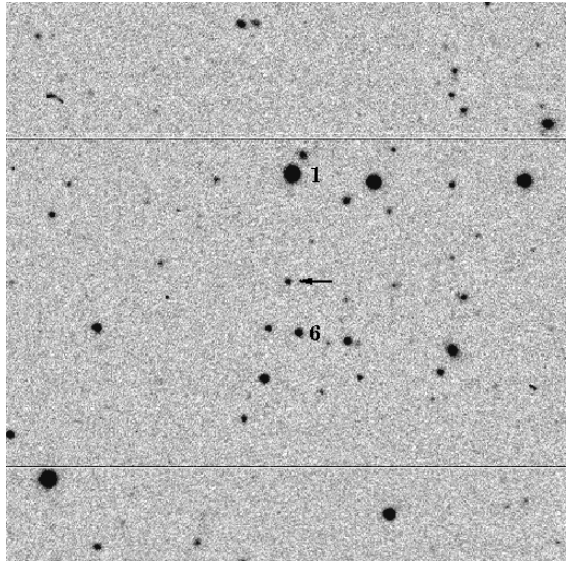}
\figcaption{QUEST image from 3.28 hours after the burst.
This image was taken in the drift scan mode starting 3.28 hours after the burst
, and is a combination of the two R-band images with a limiting magnitude of 
roughly $R = 22.1$ at the $S/N = 3$ level. North is up, east is to the left, 
and the image shows an area 
8' by 8'.  The optical transient is indicated with an arrow, while the 
comparison stars \#1 and \#6 are labeled.}
}
\end{figure}

\begin{figure}
{\epsscale{0.4}
\plotone{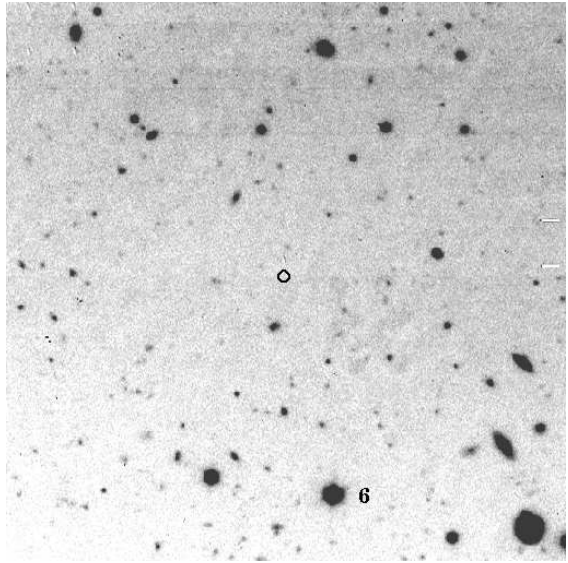}
\figcaption{Keck R-band image from 103 days after the burst.
The position of the optical transient (indicated by the open circle) is empty, 
with the nearest source being an $R = 24.7$ galaxy 6.3'' distant.  The
limiting magnitude is $R = 25.7$ at the $S/N = 3$ level, north is up, east
is to the left, and the image shows a square area 2' on a side. Our comparison
 star \#6 is labelled near the bottom.}
}
\end{figure}


\begin{references}

\reference{akerlof} Akerlof, C. et al. 1999, Nature, 398, 400

\reference{barthelmy} Barthelmy, S. et al. 1994, Gamma-Ray Bursts, eds G.
Fishman, J. Brainerd, and K. Hurley (New York, AIP, conf. proc. 307), 643

\reference{bessel} Bessel, M. S. 1990, PASP, 102, 1181 

\reference{bloom} Bloom, J. S., Sigurdsson, S., \& Pols, O. R. 1999,
MNRAS, 305, 763

\reference{castro} Castro-Tirado, A. J. et al. 1998, Science, 279, 1011

\reference{costa} Costa, E. et al. 1997, Nature, 387, 783

\reference{fenimore01} Fenimore, E., Epstein, R., Ho, C., Klebesadel,
R., \& Laros, J. 1992, Nature, 357, 140

\reference{fenimore02} Fenimore, E. E. et al. 1993, Nature, 366, 40

\reference{frail} Frail, D. et al. 1998, Nature, 395, 663

\reference{galama01} Galama, T. J. et al. 1998, ApJ, 500, L97

\reference{galama02} Galama, T. J. et al. 1999, Nature, 398, 394

\reference{harrison} Harrison, F. A. et al. 1999, ApJLett, submitted
(astro-ph/9905306)

\reference{hurley01} Hurley, K. et al. 1992, A\&A Suppl., 92, 401

\reference{hurley02} Hurley, K. et al. 1999, ApJSupp, 120, 399

\reference{kulkarni} Kulkarni, S. R. et al. 1998, Nature, 395, 663

\reference{landolt} Landolt, A. 1992, AJ, 104, 340

\reference{levine} Levine, A., Bradt, H., Cui, W., Jernigan, J., Morgan,
E., Remillard, R., Shirey, R., \& Smith, D. 1996, ApJ, 469, L33

\reference{matthews}Matthews, W. \& Soifer, B. 1994, in Infrared
Astronomy with Arrays, ed. I. McLean (Kluwer), p. 239

\reference{mendez} Mendez, R., Depoy, D., \& Bailyn, C. 1998, NOAO
News., 55, 17

\reference{monet} Monet, D. et al. 1998, USNO-A2.0 Catalog,

\reference{http}http://www.nofs.navy.mil/projects/pmm/catalogs.html

\reference{oke} Oke, B. et al. 1995, PASP, 107, 375

\reference{paczynski} Paczynski, B. 1998, ApJ, 494, L45

\reference{park01} Park, H.-S. et al. 1997, ApJ, 490, L21

\reference{park02} Park, H.-S. et al. 1998, in Gamma-Ray Bursts, eds C.
Meegan, R. Preece, and T. Koshut (New York, AIP, conf. proc. 428), 842

\reference{rocca} Rocca-Volmerange, B. \& Guideroni, G. 1988, A\&AS, 75,
93

\reference{sabbey} Sabbey, C. N., Coppi, P., \& Oemler, A. 1998, PASP, 110,
1067

\reference{sari} Sari, R., Piran, T., \& Narayan, R. 1998, ApJ, 497, L17

\reference{schaefer} Schaefer, B. E. 1999, ApJ, 511, L79

\reference{schlegel} Schlegel, D. J., Finkbeiner, D. P., \& Davis, M. 1998,
ApJ, 500, 525

\reference{smith01} Smith, D. A. et al. 1999a, ApJ, submitted

\reference{smith02} Smith, D. A. et al. 1999b, GCN no. 275

\reference{snyder} Snyder, J. A. 1998, Proc. SPIE, 3355, 635

\reference{totani} Totani, T. 1997, ApJ, 486, L71

\reference{paradijs} van Paradijs, J. et al. 1997, Nature, 386, 686

\end{references}
\end{document}